\documentclass[11pt]{article}\usepackage[dvips]{graphicx}
\usepackage{latexsym}
\usepackage{amssymb}

\newcommand{\RP}{\mathbb{RP}}
\newcommand{\R}{\mathbb{R}}
\usepackage{amsmath, amsthm, amssymb}
\newtheorem{theorem}{Theorem}

\def\bea{\begin{eqnarray}}
\def\eea{\end{eqnarray}}

\def\be{\begin{equation}}
\def\ee{\end{equation}}
\def\theequation{\thesection.\arabic{equation}}
\def\p{\partial}

\begin{document}
\title{\vskip -70pt
\begin{flushright}
{\normalsize DAMTP-2008-58} \\
\end{flushright}
\vskip 80pt
{\bf Cosmic Jerk, Snap and Beyond}
\vskip 20pt}
\author{Maciej Dunajski\thanks{\tt m.dunajski@damtp.cam.ac.uk}, 
Gary Gibbons\thanks{\tt G.W.Gibbons@damtp.cam.ac.uk}\\[8pt]
{\sl Department of Applied Mathematics and Theoretical Physics} \\
{\sl University of Cambridge} \\
{\sl Wilberforce Road, Cambridge CB3 0WA, UK} \\[8pt]}
\date{} 
\maketitle
\begin{abstract}
{We clarify the procedure for expressing the Friedmann equation 
in terms of directly measurable cosmological scalars constructed out of
higher derivatives of the scale factor. We carry out this procedure for 
pure dust, Chaplygin gas and generalised Chaplygin gas 
energy--momentum tensors. In each case it leads to a constraint on the scalars
thus giving rise to a test of General Relativity.
We also discuss a formulation of the Friedmann
equation as unparametrised geodesic motion and its connection with
the Lagrangian treatment of perfect fluids coupled to gravity.} 
\end{abstract}
\section{Introduction}
\setcounter{equation}{0}
It is a striking  and slightly puzzling fact that almost all current
cosmological observations
can be summarised by the simple  statement \cite{j2,j3,statefinder_cite}
\begin{quote} {The jerk of the universe equals  one.}\end{quote}
The aim of this paper is to uncover the mathematical
reason behind this fact and explore some of its consequences,
and  generalisations.

Our present universe appears to be well 
described by the Friedmann--Lemaitre--Robertson--Walker (FLRW) metric
\be
\label{FLRW}
g=-dt^2+a(t)^2 h,
\ee
where the function $a=a(t)$ is the scale factor, 
$h$ is a metric on $H^3$, $\R^3$ or $S^3$ with constant 
curvature $k=-1, 0$ or $1$ respectively and the speed of light has been 
set to 1.

One may define four cosmological scalars  by 
\be
\label{scalars}
H=\frac{1}{a}\frac{d a}{d t}, 
\quad q=-{a}
\Big(\frac{d a}{d t}\Big)^{-2}\frac{d^2 a}{d t^2}, \quad
Q=a^2
(\frac{d a}{d t}\Big)^{-3}\frac{d^3 a}{d t^3}, 
\quad 
X=a^3
\Big(\frac{d a}{d t}\Big)^{-4}\frac{d^4 a}{d t^4}.
\ee
These scalars are known as
Hubble, deceleration, jerk and snap respectively\footnote{
 The analogous expressions involving 5th and 6th derivatives of $a$ are
known as crackle and pop.
This terminology goes back to a 1932 advertisement of
Kellogg's Rice Crispies which   `merrily snap, crackle, and pop in a bowl of milk'. 
}. The last three of these are
dimensionless and the Hubble has the dimension of inverse of time. 
The scalars can in principle be measured using 
 red-shift data and the Hubble law \cite{Harrison, Hut, j1, 
statefinder_cite, j2, j3,  Visser}.
They  arise, for example,  as the first four terms in the Taylor expansion
of the red--shift $1+z= a(t_o)/a(t_e)$ observed at time $t=t_o$ 
 as a function of time of emission $t=t_e$.  
It is important to stress that the Einstein equations have
played no  role so far. The form 
of the FLRW metric is derived from  symmetry consideration
which requires no knowledge of the dynamical variable
$a(t)$. The Hubble law is an observational fact also independent of
the field equations. 

In this paper we shall regard  the 
Einstein equations, in the form of Friedmann equation
for $a=a(t)$, as one algebraic
constraint between the scalars (\ref{scalars}). This links the measurement
of the cosmological scalars to a test of General Relativity,
or any of its modifications (which would lead to different constraints).
If one {\it assumes} that Einstein equations hold then measuring
the cosmological scalars could determine the the equation of state in relating
the energy density and the momentum in the perfect fluid energy momentum 
tensor.
We aim to clarify this procedure (which appears to have been initiated
in \cite{Harrison} and developed by \cite{Hut} and 
is closely related  to the Statefinder
approach of \cite{statefinder_cite}) and link it to some recent work 
\cite{BDE08} 
on the metrisability of projective structures.
For example it turns out that the constant jerk condition
$Q=1$
is equivalent to Einstein equations with $k=0$ and the dust
stress tensor. Using the scaling and translational symmetries
of the ODE $Q=1$ allows to put the general solution in the form
\[
a(t)=\sinh^{\frac{2}{3}}
{\Big(\sqrt{\frac{3\Lambda}{4}t}\Big)},
\]
where $\Lambda$ is the cosmological constant.
Another motivation comes from quantum cosmology.
In the Hamiltonian treatment of general relativity one
regards the Einstein equation as a dynamical system
equivalent to a forced geodesic motion on
an infinite dimensional space ${\cal M}$
of three--dimensional 
Riemannian metrics on initial data surface $\Sigma$.
The metric on 
${\cal M}$ is the DeWitt metric
\[
{\cal G}_g(\delta g, \delta g)=\int_{\Sigma} \sqrt{g}(\mbox{Tr}(\delta g^2)- 
\mbox{Tr}(\delta g)^2) d^3 x 
\]
(where $\delta g\in T_g{\cal M}$ and the traces are taken with respect to $g$)
and the potential is given by the scalar curvature of the 
Riemannian metric on $\Sigma$. See e.g. \cite{FM71} where this is discussed. 
In the context of quantum cosmology 
this analysis needs to be modified in the presence of matter and
cosmological constant. 
In Section \ref{sec_3} we shall use a different point of view to relate 
the Friedmann equation a geodesic motion on
two--dimensional surface of revolution. In Section \ref{sec_4} we
shall make contact with the Lagrangian treatment 
of perfect fluids coupled to gravity. The problem
of finding a metric whose unparametrised geodesics coincide with 
the integral curves of  a second order ODE is summarised in the Appendix. 
\section{Cosmological scalars and the Friedmann equation}
Consider the  Einstein equations with cosmological
constant $\Lambda$ and matter  
described by the perfect fluid energy momentum 
tensor
\be
\label{fluid_tensor}
T_{\mu\nu}=(\rho+p) U_\mu U_\nu+p g_{\mu\nu}
\ee
where $U_\mu=\nabla_\mu t$, $\rho=\rho(t)$ describes the matter density
and $p=p(t)$ is the pressure.
The energy--momentum conservation yields 
\[
a\frac{d \rho}{d t}+3(\rho+p)\frac{d a}{d t}=0
\]
and the Einstein equations reduce to the Friedmann equation
\be
\label{Friedmann}
\Big(\frac{d a}{d t}\Big)^2+k=\frac{8\pi G}{3}\rho a^2+\frac{\Lambda}{3}a^2.
\ee
\vskip5pt
{\bf Pure dust.} Consider the pressure--free
matter $p=0$. The energy conservation gives  
$\rho a^3=M$ where $M$ is a constant. 
Let us now consider a system of three equations
consisting of (\ref{Friedmann}) and its first two time derivatives.
We regard this as a system of algebraic equations for
the constants $(k, \Lambda, GM)$ which can therefore be expressed
as functions of $(a, \dot{a}, \ddot{a}, \dddot{a})$. Take
the third derivative of (\ref{Friedmann}) and substitute the
expressions for $(k, \Lambda, GM)$. 
The resulting equation\footnote{This idea of eliminating parameters and
reinterpreting them as constants of integration
has a long history. It goes back at least to 
Halphen \cite{Halphen} who obtained
a fifth  order ODE characterising conics in $\RP^2$. 
In the inhomogeneous coordinates $(a, t)$ 
the five parameter family of conics  is
\[
a^2=c_1\;t^2+c_2\;at+c_3\;a+ c_4\;t+c_5.
\]
Eliminating the parameters $(c_1, \dots, c_5)$ between this equation
and its fourth derivatives and substituting in the fifth derivative
yields the ODE
\[
\frac{d^3}{d t^3}\Big(\Big(\frac{ d^2 a}{d t^2}\Big)^{-\frac{2}{3}}\Big)=0.
\]
}
does not contain any parameters and
can be expressed
in terms of the cosmological scalars (\ref{scalars}) as
\be
\label{our_constraints}
X+2(q+Q)+qQ=0.
\ee
This fourth order ODE is equivalent to the Friedmann equation
and has an advantage that it appears as a constraint on directly
measurable quantities. Thus it provides the test of the model
alluded to in the Introduction. If only two constants
$(\Lambda, GM)$ are eliminated between (\ref{Friedmann}) and its first
derivative then the second derivative of  (\ref{Friedmann}) yields
\[
k=a^2H^2(Q-1)
\]
where $k$ is regarded as a parameter. This is the formula
obtained in \cite{Harrison}. In particular if $k=0$ this relation reduces
to a  third order ODE 
\[
Q=1.
\]
This constant
jerk condition is consistent with recent redshift
analysis which provides the lower bound for the jerk implying that
$Q>0$. Compare a related discussion in \cite{Visser}.
{\vskip 5pt} {\bf Chaplygin Gas.} This exotic form of matter
is given by the perfect fluid energy momentum tensor with the equation of 
state
\[
p=-\frac{A}{\rho},
\]
where $A$ is a positive constant. 
In the cosmological context
this model was proposed in \cite{chaplygin_cite}.
The negative pressure allows to describe a
transition from a decelerated universe to cosmic acceleration. 
The energy conservation equation gives
\[
\rho=\sqrt{A+\frac{B}{a^6}}. 
\]
For small $a(t)$ this reduces to  the dust cosmology 
$\rho= \sqrt{B}a^{-3}$, and for large $a$  one gets
the de Sitter Universe  $\rho =\sqrt{A}, p=-\sqrt{A}$. In between these 
two regimes one can use the expanded expression
\[
\rho =\sqrt{A}+\frac{B}{2\sqrt{A}} a^{-6}. 
\]
Thus $\sqrt{A}$ plays the role of a cosmological constant.
We insert this to the Friedmann equation with $\Lambda =0$ 
and follow the procedure
of eliminating the constants by differentiation. This leads
to an approximate constraint
\be\label{our_constraints2}
X+5(Q+q)+qQ=0
\ee
which is valid when $a^{-6}B/2\sqrt{A}$ is small compared to the unity.

{\vskip 5pt} {\bf Generalised Chaplygin Gas.} This is the generalisation
of the previous case, where
\[
p=-\frac{A}{\rho^{\alpha}}, \qquad 0\leq \alpha\leq 1.
\]
Introducing an additional constant $\alpha$ allows a description of a universe
evolving from the pure dust matter to a cosmological constant with an 
intermediate epoch with non--zero cosmological constant and
perfect fluid matter satisfying the equation of state $p=\alpha \rho$.

The energy conservation gives
\[
\rho=(A+Ba^{-3(\alpha+1)})^{\frac{1}{\alpha+1}}\approx
A^{\frac{1}{\alpha+1}}+\frac{A^{\frac{1}{\alpha+1}}}{\alpha+1}
\frac{B}{A}a^{-3(\alpha+1)}+\dots\,\,.
\]
Inserting the expanded formula
into the Friedmann equation with $\Lambda=0$ 
leads to an expression with three arbitrary 
constants. Our procedure gives the constraint
\[
X+(3\alpha+2)(Q+q)+qQ=0
\]
in agreement with (\ref{our_constraints}) and (\ref{our_constraints2})
when $\alpha=0$ and $\alpha=1$ respectively.
We may however want to eliminate $\alpha$ from this equation by another
differentiation. This requires an additional cosmological scalar
crackle
\[
Y=a^4\Big(\frac{d a}{d t}\Big)^{-5}\frac{d^5 a}{d t^5}.
\]
A rather complicated (MAPLE) calculation leads to a relatively simple 
constraint
\[
-2qX-2Q{q}^{2}-Yq-2XQ-3X{q}^{2}-{Q}^{2}q-YQ+{X}^{2}-3{q}^{3}
Q-qXQ+{Q}^{3}-2{Q}^{2}{q}^{2}=0
\]
which should hold if Einstein equations with the 
generalised Chaplygin gas energy momentum holds. 
This constraint is again approximate and is valid only
in the regime where the higher order terms in the expansion of $\rho$ can 
be dropped.

A different constraint involving $Y$ appeared in the work of Hut 
\cite{Hut} who considered a mixture of perfect dust and radiation in the energy momentum tensor.
\section{Friedmann equations as geodesic motion}
\label{sec_3}
In this section we shall relate the Friedmann equation to a 
geodesic motion on a
surface of revolution with metric given explicitly by (\ref{mini_dewit}).

For simplicity we consider the pure dust case $p=0$. 
Eliminating the constant $GM$ 
between the Friedmann equation and its time derivative 
yields a second order ODE
\be
\label{2nd_Friedmann}
\frac{d^2 a}{d t^2}=\frac{1}{2}\Big(\Lambda a -\frac{k}{a}\Big)-\frac{1}{2a}\Big(\frac{d a}{d t}\Big)^2.
\ee
Can we think of the integral curves of this equation as unparametrised geodesics of some metric
on a surface with local coordinates $(a, t)$? This question can be asked about any second order 
ODE of the form $\ddot{a}=F(t, a, \dot{a})$. The necessary
condition is that $F$ is at most cubic in the first derivatives $\dot{a}$. This condition (which was
known to R. Liouville \cite{Liouville}) is obviously 
satisfied by (\ref{2nd_Friedmann}). Therefore the integral curves
of (\ref{2nd_Friedmann}) are geodesics of some projective structure 
(an equivalence class of torsion--free connections
sharing the same unparametrised geodesics).
In \cite{Liouville} Liouville  also began the study of sufficient conditions for a projective structure to come from a metric, 
but these were derived only recently \cite{BDE08}. They come down to vanishing of
three weighted invariants of differential order at most 8 in connection defining the projective structure. 
We have checked that these invariants vanish of (\ref{2nd_Friedmann}) and so there exist an underlying metric.
Following a linear algorithm lied down in \cite{Liouville} 
and summarised in the Appendix
one can construct the metric
explicitly. The answer is
\be
\label{mini_dewit}
{\cal G}=\frac{dt^2}{ka-\frac{\Lambda}{3} a^3+c}
+\frac{a\,da^2}{(ka-\frac{\Lambda}{3} a^3+c)^2},
\ee
where $c$ is a constant equal to a negative multiple of the Newton's
constant.
To verify this, write the geodesic equations for (\ref{mini_dewit})
and eliminate the affine parameter between the two equations thus
expressing $a$ as a function of $t$. This will lead back to 
(\ref{2nd_Friedmann}).
The scalar curvature of the metric (\ref{mini_dewit}) is
\[
\frac{2k\Lambda a^3+9\Lambda a^2c+6 k^2a+3k c}{6a^2}.
\]
Thus the metric has constant curvature (and therefore it is 
projectively flat) if and only if $k=0$. In this case there exist a coordinate transformation $(t, a)\rightarrow (\hat{t}(\textit{}t, a), \hat{a}(t, a))$ 
such that, in the new coordinates
the Friedmann equations are
\[
\frac{d^2 \hat{a}}{d\hat{t}^2}=0.
\]
The cubic denominators in  (\ref{mini_dewit}) have real zeroes,
and so the metric may appear singular. In fact these zeroes correspond 
to coordinate singularities, and are not present in curvature scalars.
For example when $k=0$ the curves $a^3=3c/\Lambda$ are only apparent
singularities as the metric has constant curvature.
\section{Lagrangian description of perfect fluids}
\label{sec_4}
In this section we shall make contact with a standard 
Lagrangian treatment of perfect fluids coupled to 
gravity \cite{GT08} and then specialise to the FLRW case. 
We shall then use this formalism to 
derive the  geodesic  formulation analogous to (\ref{mini_dewit}) when radiation and other forms of matter are present in the energy--momentum tensor.

We shall restrict our general discussion to the case of an irrotational (i. e. vorticity--free) perfect fluid. We introduce a potential $\psi$  and let 
\[
Z=-g^{\mu \nu} \nabla_\mu \psi \nabla _\nu \psi \,
\]
We take  as  Lagrangian for $\psi$   
\[
L=L(Z)\,.
\]
The energy momentum tensor is 
\[
T_{\mu \nu} = 2  L_Z\nabla_\mu \psi \nabla _\nu \psi + L g_{\mu \nu}.
\]
Comparing this with  a perfect fluid (\ref{fluid_tensor})
with $U_{\mu}=(\nabla_\mu \psi)/Z$
we may identify
\[
\rho= 2ZL_Z-L \,\qquad p= L\,.
\]
Adding a constant to the Lagrangian $L(Z)$ introduces
the cosmological term in the FLRW models:
\[
L\rightarrow L+\gamma\quad \rho\rightarrow \rho-\gamma,\quad
p\rightarrow p+\gamma, \quad T_{\mu\nu}\longrightarrow T_{\mu\nu}
+\gamma g_{\mu\nu}.
\]
\begin{itemize}
\item If  $p={\alpha\rho} $ then 
$L= Z^{\frac{1+\alpha}{2\alpha}}$.
\item For a Born-Infeld scalar field 
\[
L= -\sqrt{1-Z} +1 
\]
we get
\[
p= \frac{\rho}{1+ \rho}.
\]
Omitting the one, which amounts to changing the zero of the energy
 scale, or cancelling a cosmological term,  gives the Chaplygin gas
\[
p=-\frac{1}{\rho}
\]
mentioned earlier.
\end{itemize}
The shift symmetry $\psi \rightarrow \psi +{\rm constant}$ 
gives rise to a conserved current  which, in this perfect  fluid context, may be interpreted as the entropy current.
\[
s^\mu =-\frac{\p L}{\p ( \nabla_\mu \psi ) } \,,\qquad \nabla _\mu s ^\mu =0\,.
\] 
As an example, consider radiation $L=Z^2$.
The equations of motion are
\be
\label{conf_eq}
\nabla _\mu \bigl 
( ( \nabla \psi) ^2    \nabla ^\mu \psi \bigr ) =0\, ,  
\ee
or, as long as $\nabla ^\mu  \psi$ is time-like 
\[
\bigl ( g^{\mu \nu} - 2 U^\mu U^\nu \bigr) \nabla _\mu \nabla _\nu \psi =0\,. 
\]
Two simple solutions in flat space-time are
\begin{itemize} 
\item
$\psi=f(t-z)$, where $z$ is one of the spatial coordinates
and $f$ is an arbitrary function.
This gives a trivial solution with vanishing energy momentum tensor.
\item
$\psi =t$. This represents a uniform fluid at rest.
Linear perturbations about this, or indeed any,  solution
are easily seen to travel with respect to it with speed 
$\frac{1}{\sqrt{3}}$.  
\end{itemize} 
Interestingly  the action, and hence the equations of motion, are invariant 
under Weyl conformal rescalings  
\[
\psi \rightarrow \psi\,,\qquad
 g_{\mu \nu} \rightarrow \Omega ^2(x) g_{\mu \nu}\,.
\]  
In other words the equations (\ref{conf_eq}) are conformally invariant, 
just as is  Yang--Mills theory in four space-time dimensions.
\subsection{Coupling to FLRW models}
To couple to gravity we would consider the Lagrangian
( up to a boundary term)
\be
\int \Bigl ( \frac{R}{16 \pi G} + L(Z) \Bigr) \sqrt {-g} d^4 x. 
\label{action}   
\ee
To obtain the full consequences of the 
Einstein equations we substitute the ansatz $\psi=\psi(t)$ and 
\be
\label{FLRW2}
g=-N^2 dt^2+a(t)^2 h,
\ee
where $N=N(t)$ is an arbitrary lapse function
into the action (\ref{action}). 
We vary with respect to  $\psi, a $ and $N$.
The first two Euler-Lagrange 
equations are second order in time.  The third  variation
gives a constraint on the initial values $\psi, \dot \psi $ and $a,
\dot a $
which is consistently evolved by the two second order equations.
It is convenient to make the choice  $N=1$ after variation.
It then follows that the constraint is just the vanishing of the
Hamiltonian $\cal H$ 
obtained from the Lagrangian 
\be
\label{effective_lagrangian}
{\cal L}= \frac{a^3}{ 4}L(Z)-
\frac{3}{8 \pi G}(a (\dot a)^2 -ka),\quad\mbox{where}\quad  Z=\dot{\psi}^2
\ee
by a standard Legendre transform.
One may  check that the resulting system of ordinary differential  
equations is the same as the system one would obtain
by substitution of the ansatz (\ref{FLRW2}) into  the full Einstein 
equations.  

Clearly, for a general equation of state $p=p(\rho)$,
the kinetic term in the Lagrangian
for  $\psi$ will not be quadratic in $\dot \psi$ 
and hence the standard Jacobi procedure will not lead to a metric.
The exceptional case is of course `stiff matter', $p=\rho$ 
which is well known to be represented by a standard massless scalar
field. We shall instead use the procedure introduced in 
Section \ref{sec_3} to construct a metric whose geodesics coincide with
unparametrised extremals of the Euler--Lagrange equations
of (\ref{effective_lagrangian}).

The conservation of momentum $a^3\dot{\psi}L'(Z)=$const 
can be used to find $\dot{\psi}=\dot{\psi}(a)$. It is convenient
to introduce a function $f=f(a)$ such that
\[
\frac{d f}{d a}=2\pi G a^2 L(Z(a))+k.
\]
The equations
of motion resulting from (\ref{effective_lagrangian})
become
\be
\label{second_ODE}
\frac{d^2 a}{d t^2}=-\frac{1}{2a}\,
\frac{d f}{da}-\frac{1}{2a}\Big(\frac{d a}{ d t}\Big)^2.
\ee
Examining the obstructions of \cite{BDE08} shows that
the projective structure defined by (\ref{second_ODE}) is
metrisable for any $f(a)$ (and therefore for any choice of the Lagrangian $L$). 
Following the algorithm of \cite{BDE08}
leads to the expression for the metric
\be
\label{metric_2}
{\cal G}=\frac{d t^2}{f(a)}+\frac{ a\,da^2}{f(a)^2}.
\ee
Unparametrised geodesics of this metric are integral curves
of (\ref{second_ODE}). For example if $p=\alpha\rho$ and the cosmological constant is allowed
then  $L(Z)=Z^{\frac{1+\alpha}{2\alpha}}+\gamma $ and
\[
f(a)=-\frac{2\pi\alpha G}{3\alpha+1}\Big(\frac{b}{1+\alpha}\Big)^{1+\alpha}
a^{-3\alpha-1}
+\frac{2\pi G\gamma}{3} a^3+k a +c.
\]
where $b$ and $c$ are some constants. The pure dust case
considered in Section \ref{sec_3} is recovered in the limit
$\alpha\rightarrow 0$. In this case $f$ is cubic in $a$ and
the metric (\ref{metric_2}) becomes (\ref{mini_dewit}) 
if $\gamma=-\Lambda/(2\pi G)$.

For  radiation $(\alpha=1/3)$  the Legendre transform gives the Hamiltonian and the momenta
\begin{eqnarray*}
{\cal H} &=& \frac{ 3  a^3}{4} (\dot \psi)^4 -\frac{ 3}{8 \pi G} 
(a (\dot a)^2 +ka )\,,\\
p_\psi &=& a^3 (\dot \psi)^3\,,\\
p_a &=& -\frac{ 3 }{ 4 \pi G} a^2 H, 
\end{eqnarray*}
where $H=\dot{a}/a$. Note that the conserved momentum $p_\psi$ is essentially the conserved entropy of the radiation (i.e. number of photons). The phase $\psi$ is  ignorable both in the technical and the 
vernacular sense since  it disappears from the Hamiltonian.  
\section{Summary}
We have studied the Friedmann equation from the point of view of geometry
of ODEs. 

In Section 2 we set up an algorithm  replacing 
the 2nd order Friedmann equation which involves
a number of parameters (this number depends on a choice
of equation of state) by a constraint on directly observable
cosmological scalars. The validity of resulting constraints can in 
principle be verified by observations, thus providing the experimental
test of General Relativity as well as rulling out some equations of state.
For example the Friedmann equation with $k=0$ and the dust energy--momentum tensor is equivalent to the constant jerk condition $Q=1$.

The constraints discussed in Section 2 give rise to curves
in the $(a, t)$ plane, one curve through each point in each direction, 
and in Sections 3 and 4 we have demonstrated that these curves
are unparametrised geodesics
of some two--dimensional (pseudo)--Riemannian metric. This is an alternative
to the more usual DeWitt approach and may play a role
in projective quantisation of Cosmology which does not refer
to a specific (time) parametrisation of classical trajectories.
\section*{Appendix}
\setcounter{equation}{0}
\appendix
\def\theequation{\thesection{A}\arabic{equation}}
Recall that a projective structure 
on an open set $U\subset \R^2$ is an equivalence class
of torsion free connections on $TU$. 
Two connections  are projectively equivalent if they share the same unparametrised geodesics. Let $(a, t)$ be local coordinates
on $U$. Eliminating a parameter $s$ between the two geodesic
equations 
$
\ddot{x}^i+\Gamma^i_{jk}\dot{x}^j\dot{x}^k=v\dot{x}^i
$
where $x^i(s)= (t(s), a(s))$ yields a second order ODE
\begin{equation}
\label{ODE2}
\frac{d^2 a}{d t^2}=A_3(t, a)\Big(\frac{d a}{d t}\Big)^3
+A_2(t, a)\Big(\frac{d a}{d t}\Big)^2
+A_1(t, a)\Big(\frac{d a}{d t}\Big)+A_0(t, a)
\end{equation}
where the functions $A_\alpha$ are given in terms of the
coefficients of a connection in a given projective class
\be
\label{expressions_forA}
A_0=-\Gamma^2_{11},\quad A_1=\Gamma^1_{11}-2\Gamma^2_{12}, \quad 
A_2=2\Gamma^1_{12}-\Gamma^2_{22},\quad A_3=\Gamma^1_{22}.
\ee
Conversely, any ODE of the form (\ref{ODE2}) gives
rise to a projective structure.

 A projective structure is called
metrisable if the integral curves of the ODE (\ref{ODE2})
are unparametrised geodesics of a metric
\begin{equation}
\label{metric}
g=E(t, a)dt^2+2F(t, a) dtda+G(t, a)da^2.
\end{equation}
In this case the four functions $A_\alpha$ are given in
terms of $(E, F, G)$ and their first derivatives.
The corresponding expressions can be derived
by calculating the Levi--Civita connection of $g$ and
reading off the $A_\alpha$s from (\ref{expressions_forA}).
The substitution
\[
E=\psi_1/\Delta^2, \quad F=\psi_2/\Delta^2, \quad
G=\psi_3/\Delta^2, \qquad \Delta =\psi_1\psi_3-{\psi_2}^2
\]
linearises these 
expressions thus leading to a characterisation of
the metricity condition as consistency conditions for
an overdetermined system of linear PDEs
\begin{theorem}[R. Liouville 1889 \cite{Liouville}]
\nonumber
A projective structure  corresponding
to the second order ODE {\em(\ref{ODE2})}
is metrisable on a neighbourhood of a point $p\in U$ iff there exists functions
$\psi_1, \psi_2, \psi_3$ defined on a neighbourhood of $p$ such that
\[
\psi_1\psi_3-{\psi_2}^2
\] 
does not vanish at $p$ and such that the equations
\begin{eqnarray}
\label{linear_system}
\frac{\p\psi_1}{\p t}&=&\frac{2}{3}A_1\psi_1-2A_0\psi_2,\nonumber\\
\frac{\p \psi_3}{\p a}&=&2A_3\psi_2-\frac{2}{3}A_2\psi_3,\nonumber\\
\frac{\p \psi_1}{\p a}+2\frac{\p \psi_2}{\p t}&=&\frac{4}{3}A_2\psi_1
-\frac{2}{3}A_1\psi_2-2A_0\psi_3,\nonumber\\
\frac{\p \psi_3}{\p t}+2\frac{\p \psi_2}{\p a}&=&
2A_3\psi_1
-\frac{4}{3}A_1\psi_3+\frac{2}{3}A_2\psi_2
\end{eqnarray}
hold on the domain of definition.
\end{theorem}
This system 
is overdetermined, as there are more
equations than unknowns. In \cite{BDE08} the consistency
conditions for this system where
found in terms of point invariants of the associated
second order ODE (\ref{ODE2}). The details of this construction
and the invariants themselves are rather complicated
and we refer the reader to \cite{BDE08}. 
The invariants obstructing metricity vanish for the projective structures (\ref{2nd_Friedmann}) and (\ref{second_ODE}) arising in our paper, and the corresponding metrics
are found by solving the linear system 
(\ref{linear_system}).

\end{document}